\documentclass[conference]{IEEEtran}
\IEEEoverridecommandlockouts

\usepackage{cite}
\usepackage{amsmath,amssymb,amsfonts}
\usepackage{algorithmic}
\usepackage{graphicx}
\usepackage{textcomp}
\usepackage{xcolor}
\usepackage{subcaption}
\def\BibTeX{{\rm B\kern-.05em{\sc i\kern-.025em b}\kern-.08em
    T\kern-.1667em\lower.7ex\hbox{E}\kern-.125emX}}

\newcommand{\ua}{\uparrow}
\newcommand{\nc}{\newcommand}
\nc{\da}{\downarrow} \nc{\hc}{\hat{c}} \nc{\hS}{\hat{S}}
\nc{\bra}{\langle} \nc{\ket}{\rangle}
\nc{\h}{\hat} \nc{\hT}{\h{T}}\nc{\rd}{\textrm{d}}\nc{\e}{eqnarray}\nc{\hR}{\hat{R}}\nc{\Tr}{\mathrm{Tr}}
\nc{\tS}{\tilde{S}}\nc{\tr}{\mathrm{tr}}\nc{\8}{\infty}\nc{\lgs}{\bra\ua,\theta|}\nc{\rgs}{|\ua,\theta\ket}
\nc{\hU}{\hat{U}}\nc{\lfs}{\bra\theta|}\nc{\rfs}{|\theta\ket}\nc{\hZ}{\hat{Z}}\nc{\hd}{\hat{d}}\nc{\mD}{\mathcal{D}}
\nc{\bd}{\bar{d}}\nc{\bc}{\bar{c}}\nc{\mc}{\mathcal}\nc{\ea}{eqnarray}\nc{\mG}{\mathcal{G}}\nc{\bce}{\begin{center}}
\nc{\ece}{\end{center}}

\renewcommand{\vec}[1]{\ensuremath{\boldsymbol{#1}}}

\newcommand{\kron}{\otimes}

\renewcommand{\eqref}[1]{(\ref{eq:#1})}

\newcommand{\figref}[1]{Fig.~\ref{fig:#1}}
\newcommand{\tabref}[1]{Table~\ref{tab:#1}}
\newcommand{\secref}[1]{Section~\ref{sec:#1}}

\newcommand{\ie}{i.e.,\ }

\usepackage{graphicx}
\usepackage{authblk}
\usepackage[left=0.673in,
            right=0.673in,top=0.70in,bottom=1.1in]{geometry}
\begin{document}
\title{Low-Power Double RIS-Assisted Mobile\\ LEO Satellite Communications
\thanks{We would like to thank Kempestiftelserna, the Swedish Foundation for Strategic Research multidisciplinary research center SMART 6GSAT, and the European Project Hexa-X II (grant 101095759) for funding this work.}
}
\author[1]{\small Kunnathully Sadanandan Sanila}
\author[2]{\small Rickard Nilsson}
\author[3]{\small Emad Ibrahim}
\author[4]{\small Neelakandan Rajamohan}
\affil[1,2]{\footnotesize Department of Computer Science, Electrical and Space Engineering, Luleå University of Technology, Sweden}
{    \makeatletter
    \renewcommand\AB@affilsepx{, \protect\Affilfont}
    \makeatother
\affil[3]{\footnotesize Ericsson Research, Sweden}\affil[4]{\footnotesize School of Electrical Sciences, Indian Institute of Technology Goa, India}}
{    \makeatletter
    \renewcommand\AB@affilsepx{, \protect\Affilfont}
    \makeatother
\affil[1]{sankun@associated.ltu.se}
\affil[2]{rickard.o.nilsson@ltu.se}
 \affil[3]{emad.ibrahim@ericsson.com}
    \affil[4]{neelakandan@iitgoa.ac.in}}

\maketitle

\begin{abstract}
We propose a low-power mobile low earth orbit (LEO) satellite communication architecture, employing double reconfigurable intelligent surfaces (RIS) to enhance energy efficiency and signal performance. With a distance between RISs that satisfies the far-field requirement, this architecture positions one small RIS each in the near-field of the satellite's antenna and the user on the ground.  Moreover, we develop a path loss model for the double-RIS communication link, considering the near-field and far-field effects. Further, with the help of dual-stage beamforming, the proposed system maximizes the signal power and minimizes power consumption. Simulation results show that the proposed architecture can reduce the power consumption with 40 dB in the uplink, with a small $0.25^2$ $\text{m}^2$ RIS near the user, to communicate in energy-constrained LEO satellite communication circumstances.
\end{abstract}
\begin{IEEEkeywords}
Non-terrestrial mobile networks, Reconfigurable intelligent surfaces, Low-earth orbit satellites.
\end{IEEEkeywords}

\IEEEpeerreviewmaketitle

\section{Introduction}

The 5G terrestrial communication network has made consequential efforts to meet the critical specifications for ultra-reliable and low-latency communications (URLLC), massive machine-type communications (MTC), and enhanced mobile broadband (eMBB) \cite{Agiwal2016}. Nevertheless, because it is incredibly challenging and expensive to install terrestrial base stations and backhauls, conventional terrestrial wireless networks are still predominantly unable to deliver reliable communication coverage for geographically remote regions, including deserts, mountains, rural areas, and oceans\cite{Li2020}. Non-terrestrial network (NTN) systems were established under this limitation as a productive means of extending services over underserved or unexplored geographic areas in addition to terrestrial networks. Moreover, NTN communication offers far more versatility and a lower cost of coverage than terrestrial communication to facilitate the rapid development of smart telecommunication technologies \cite{Araniti2022}. The impending sixth-generation (6G) network is envisioned to incorporate the characteristics of terrestrial and non-terrestrial satellite networks to envision a fully connected society \cite{Geraci2023}.

Recently, the low earth orbit (LEO) satellite has enticed the most attention because of its low-altitude deployment and resulting benefits like reduced costs, smaller propagation loss, and shorter transmission delays than other higher-orbit satellites \cite{You2020}. However, the substantial path loss caused by the massive distance between the satellite and satellite users necessitates a high-power transmitter and a high-sensitivity receiver for LEO satellite communication \cite{Fang2021}. Consequently, LEO satellite communications will eventually require more expensive hardware and greater electrical power.
In order to improve the communication performance of LEO satellite links, reconfigurable intelligent surfaces (RISs) are deployed either on the satellite \cite{Tekbiyik2022} or close to ground users \cite{Dong2023}.

The contributions of the paper are summarised as follows:
\begin{itemize}
 \item \emph{Novel low-power system architecture:} The paper presents an innovative low-power mobile satellite communication architecture that maximizes the performance of the communication link by utilizing two RISs.
 \item \emph{Combined path loss analysis:} The study contributes an exhaustive path loss model by integrating both near-field and far-field models.
 \item \emph{Dual-stage beamforming:} Dual-stage beamforming is enforced in the proposed arrangement to reduce power consumption and increase signal power.
 \item \emph{Simulation and performance analysis:} The results show the DRIS-assisted configuration's capability for promoting efficient and economically viable mobile LEO satellite communication networks.
\end{itemize}

\emph{Notation}:  The symbols $\vec{a}$, $\vec{A}$, and $\mc{A}$ represent vectors, matrices, and sets, respectively. The notation $\mc{N}(\vec{\mu},\vec{C})$ refers to a Gaussian distribution with mean $\vec{\mu}$ and covariance $\vec{C}$, while $\mc{CN}(\vec{\mu},\vec{C})$ denotes a complex Gaussian distribution with the same parameters. The operations $(\cdot)^T$, $(\cdot)^*$, and $|\cdot|$ represent the transpose, Hermitian (conjugate transpose), and element-wise magnitude of a complex vector, respectively. $\mathrm{diag}(\vec{a})$ produces a diagonal matrix of $\vec{a}$. $\vec{a}[i]$ and $\vec{A}[i]$ represent the $i^{\text{th}}$ entry and $i^{\text{th}}$ diagonal entry, respectively.

\section{System Model}\label{sec:sysmodel}

\begin{figure}[htbp]
   \centering
    \includegraphics[width=0.9\linewidth]{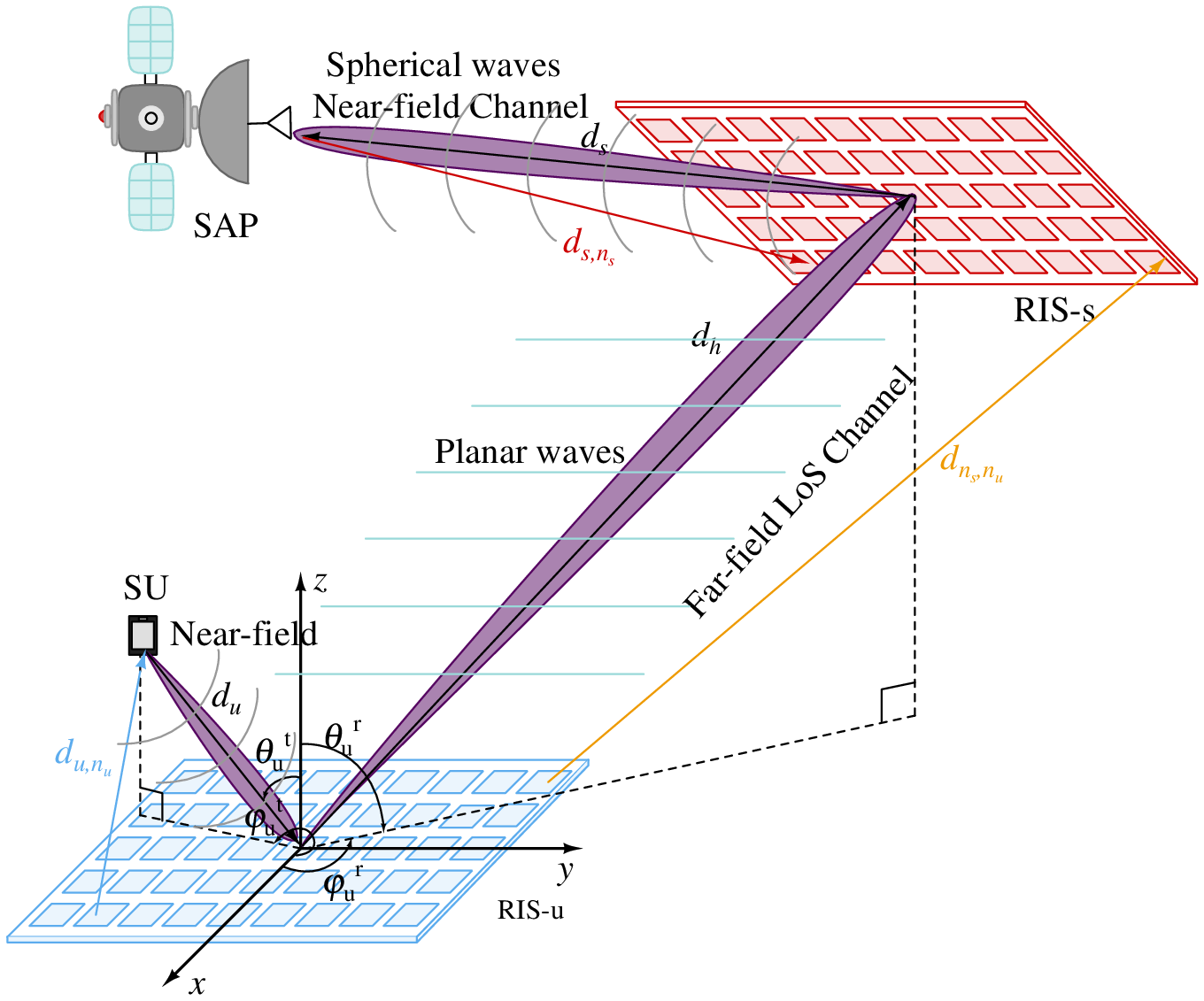}
    \caption{Double RIS architecture. RIS-s is fixed to the satellite while the mobile RIS-u is near the user on the ground.}	\label{fig:sysmodel}
\end{figure}

 In this study, we develop mobile DRIS-assisted LEO satellite communications. We consider the uplink between a single antenna satellite user (SU) (mobile) on the ground and a single antenna satellite access point (SAP) (LEO satellite) in space. {In our proposed architecture, two independent RISs are deployed in the near-field of the LEO satellite and the SU \cite{Ibrahim21} to effectively compensate for the large path losses caused by the long propagation distances, as illustrated in \figref{sysmodel}. It is assumed that the SU bears a portable RIS to enhance performance by beamforming towards the LEO satellite with a fixed RIS.} In addition, we assume that the RIS set up at the SAP has $N_{s}=N_{s,x} \times N_{s,y}$ reflecting elements (REs) that are wired to the intelligent controller. Nonetheless, those RIS that the user carries have $N_{u}=N_{u,x} \times N_{u,y}$ REs. For ease of analysis, the sets containing the REs at RIS are defined as $\mathcal{N}_i=\{1,\cdots,N_i\}$, for $i\in \{s,u\}$.
 
 We represent the baseband equivalent channels from the SU to RIS-u, RIS-u to RIS-s, and RIS-s to SAP as $\vec{h}_u \in \mathbb{C}^{N_s \times 1}$, $\vec{H} \in \mathbb{C}^{N_s \times N_u}$, and $\vec{h}_s^T \in \mathbb{C}^{1 \times N_s}$, respectively. The reflection coefficient matrix of RIS $i$ is defined as $\vec{\Phi}_i = \text{diag}\{\rho_{i,1},\cdots,\rho_{i,N_i}\}$,
where $\rho_{i,n_i}=\text{a}_{i,n_i}e^{j \phi_{i,n_i}}$. $\text{a}_{i,n_i}$ and $\phi_{i,n_i}$ denote the controllable amplitude and phase-shift of $n_i^{\text{th}}$ element $E_{i,n_i}$ at RIS $i$, for $n_i\in \mathcal{N}_i$. The controllable amplitude levels are defined as
\begin{align}\label{eq:af}
a_{i,n_i}=\begin{cases}
                  1 &\text{for passive RIS}\\
                  a &\text{for active RIS; }a> 1
                 \end{cases}.
\end{align}
We denote $L_r$ as the path loss of the double reflected path from the SAP to SU. Therefore, for the total transmit power $P_t$ from the SU, the overall received complex baseband signal at the SAP can be represented as\footnote{The direct link from the SU to the SAP is neglected due to the use of single directional antennas at both the transmitter and receiver. Additionally, the single-reflected path from SU to SAP via RIS-u is ignored because RIS-u is in the SU's near-field, making the paths from SU to SAP and RIS-u to SAP effectively similar. This approximation similarly applies to the single reflection from SU to SAP through RIS-s.}
\begin{align}\label{eq:rxsig}
 y= \sqrt{{P_t}/{L_r}}\left\{\vec{h}_s^T\vec{\Phi}_s\vec{H}\vec{\Phi}_u\vec{h}_u\right\}+z
\end{align}
and
\begin{align}
z\sim \begin{cases}
                  z_s &\text{for passive RIS-u}\\
                  \sqrt{L_s}\vec{h}_s^T\vec{\Phi}_s\vec{H}\vec{\Phi}_u\vec{w}+z_s &\text{for active RIS-u}
                 \end{cases},
\end{align}
where $z\sim\mathcal{CN}(0,N_0)$, $z_s \sim \mathcal{CN}(0,\sigma_{s}^2)$ is the static noise term and $\vec{w}  \in \mathbb{C}^{N_u \times 1} \sim \mathcal{CN}(0,\sigma_{d}^2)$ is the non-negligible additional thermal noise vector generated by power amplifiers of active REs rather than in the passive REs. Moreover, $L_s$ is the path loss from the active RIS to the SAP.
Moreover, we consider that the REs in {active RIS} have different reflection coefficients but have the same amplification factor as defined in \eqref{af} to minimize the power consumption problem of a fully connected design that requires a dedicated power amplifier for each RE \cite{Sanila23,Liu}.

\subsection{Channel Models}\label{sec:Channel}
In wireless communication systems, the electromagnetic radiation field can be classified as far-field or near-field based on the Rayleigh distance, $R_d=\frac{2D^2}{\lambda}$, where $D$ and $\lambda$ represent the array aperture and wavelength, respectively. The distinction of fields at different distances results in the formulation of varying channel models.
\subsubsection{Far-Field Planar Wave Channel Model}
We model the wireless communication channel as a far-field planar channel when the  separation connecting the transmitter and receiver is larger than the Rayleigh distance. We represent the classical far-field line-of-sight (LoS) multiple-input multiple-output (MIMO) channel between two uniform planar arrays (UPAs) as the product of the array response vectors at the transmitter UPA, $t$, and receiver UPA, $r$, nodes \cite{zheng}, \ie
\begin{align}\label{eq:ffchannel}
 \vec{H}_{r,t}^{ff}= e^{-jkd_{t,r}}\vec{a}_r(\theta_r,\varphi_r)\vec{a}_t^T(\theta_t,\varphi_t),
\end{align}
where $d_{t,r}$ is the separation between the center of the transmitter and receiver, $k=\frac{2\pi}{\lambda}$ depicts the wave number and the array response vector at node $n$, for $n = r,t$ can be defined as
\begin{align}
 \vec{a}_n&(\theta_n,\varphi_n)= \vec{\epsilon}\left(\frac{2\Delta}{\lambda}\cos(\varphi_n)\sin(\theta_n),N_{n,x}\right) \nonumber\\
 & \kron \vec{\epsilon}\left(\frac{2\Delta}{\lambda}\sin(\varphi_n)\sin(\theta_n),N_{n,y}\right) \in \mathbb{C}^{N_n \times 1},
\end{align}
where $N_n=N_{n,x} \times N_{n,y}$ is the number of antennas/elements in the UPA, $\Delta$ is the RE spacing at both nodes, $(\theta_n,\varphi_n)$ is the elevation and azimuth angle pair at the node $n$, and $\vec{\epsilon}(\vartheta, N)$ is the one-dimensional steering vector of the uniform linear array (ULA) with $\vartheta$ being the phase-shift difference of the signals from two contiguous elements and $N$ is the number of reflecting/antenna elements in the ULA. The one-dimensional steering vector can be represented as
\begin{align}
 \vec{\epsilon}(\vartheta, N)=\begin{bmatrix}
                               1 & e^{-j\pi \vartheta}&\cdots&e^{-j\pi (N-1)\vartheta}
                              \end{bmatrix}^T.
\end{align}
\subsubsection{Near-Field Spherical Wave Channel Model}
We model the multiple-input single-output (MISO) wireless communication channel as a near-field spherical channel when the distance between the transmitter and receiver is smaller than the Rayleigh distance, which can be expressed for a UPA as \cite{Cui}
\begin{align}\label{eq:nfchannel}
 \vec{h}^{nf}=  \vec{b}(\theta,\varphi,d),
\end{align}
where $\theta$ and $\varphi$ represent the elevation and azimuth angle, respectively.

The spherical wave assumption is used to derive the array steering vector for the near field channel as follows:
{
\begin{align}
 \vec{b}(\theta,\varphi,d) &=\vec{b}_x(\theta,\varphi,d)_{i_x}\kron \vec{b}_y(\theta,\varphi,d)_{i_y},\\
 [\vec{b}_x(\theta,\varphi,d)]_{i_x}&=e^{jk\left(i_x\Delta\cos(\varphi)\sin(\theta)-\frac{i_x^2 \Delta^2(1-\cos^2(\varphi)\sin^2(\theta))}{2d}\right)}\nonumber\\
 [\vec{b}_y(\theta,\varphi,d)]_{i_y}&=e^{jk\left(i_y\Delta\sin(\varphi)\sin(\theta)-\frac{i_y^2 \Delta^2(1-\sin^2(\varphi)\sin^2(\theta))}{2d}\right)}\nonumber
\end{align}}
where $i_x \in \mc{N}_{n,x}$, $i_y \in \mc{N}_{n,y}$, $k=\frac{2\pi}{\lambda}$ is the wave number, and $d$ is the distance between the center of the antenna array and the user. 

 \subsection{Double RIS-assisted Path Loss Model}
 This section develops the pass loss model for DRIS-assisted satellite communication.
 We assume that one RIS is placed in the near-field of the SAP, and the other is in the SU's near-field. At the same time, the two RIS are separated in such a way that they satisfy the far-field condition.
 The three-dimensional (3D) Cartesian coordinate locations of the SAP, SU, RIS-s, and RIS-u are denoted as $\vec{\ell}_s \in \mathbb{R}^{3 \times 1}$, $\vec{\ell}_u \in \mathbb{R}^{3 \times 1}$, $\vec{\ell}_s^r \in \mathbb{R}^{3 \times 1}$, and $\vec{\ell}_u^r \in \mathbb{R}^{3 \times 1}$, respectively.
Without loss of generality, we assume that the two RISs are equipped with UPAs. Moreover, the REs are placed along two orthogonal base directions $\vec{b}_i^{(x)} \in \mathbb{R}^{3 \times 1}$ and $\vec{b}_i^{(y)} \in \mathbb{R}^{3 \times 1}$, respectively, at RIS $i$. Thereby, the position of any particular element $n_i$ at RIS $i$ mentioning the indices in RIS $i$'s first and second base directions can be represented as $(n_i^{(x)},n_i^{(y)})$, for $n_i^{(x)}\in\{0,\cdots,N_{i,x}-1\}$ and $n_i^{(y)}\in\{0,\cdots,N_{i,y}-1\}$ with $N_i=N_{i,x}\times N_{i,y}$. The size of each RE at RIS $i$ is of subwavelength scale, which is expressed as $d_{i,x}$ and $d_{i,y}$, respectively, along the $x$ and $y$ base directions. The normalized power radiation pattern of a antenna/RE in the spherical coordinate system is
\begin{align}
 F(\theta,\varphi)=\begin{cases}
                   \cos^3\theta & \theta \in [0,\frac{\pi}{2}],\varphi \in [0,2\pi]\\
                   0 & \theta \in [\frac{\pi}{2},\pi],~\varphi \in [0,2\pi]
                  \end{cases},
\end{align}
where $\theta$ and $\varphi$ are the elevation and azimuth angles to a particular transmitting/receiving direction from the antenna/RE \cite{Tang}. The normalized power radiation pattern of the SU to RIS-u, RIS-u to SU, RIS-u to RIS-s, RIS-s to RIS-u, RIS-s to SAP and SAP to RIS-s are defined as $F_u^{tx}(\theta_{u,n_u}^{tx},\varphi_{u,n_u}^{tx})$, $F_u^t(\theta_{u,n_u}^t,\varphi_{u,n_u}^t)$, $F_u^r(\theta_{u,n_u}^r,\varphi_{u,n_u}^r)$, $F_u^{rx}(\theta_{u,n_u}^{rx},\varphi_{u,n_u}^{rx})$, $F_s^{r}(\theta_{s,n_s}^{r},\varphi_{s,n_s}^{r})$, and $F_s^{rx}(\theta_{s,n_s}^{rx},\varphi_{s,n_s}^{rx})$, respectively. The power of the incident signal into the RE $E_{u,n_u}$ can be represented as
\begin{align}
 P_{u,n_u}^{in}=\frac{P_tG_t}{4\pi d_{u,n_u}^2}F_u^{tc}E_u,
\end{align}
where $F_u^{tc}=F_u^{tx}(\theta_{u,n_u}^{tx},\varphi_{u,n_u}^{tx})F_u^t(\theta_{u,n_u}^t,\varphi_{u,n_u}^t)$, $d_{u,n_u}=\lVert \vec{\ell}_u^r+n_u^{(x)}d_{u,x}\vec{b}_u^{(x)}+n_u^{(y)}d_{u,y}\vec{b}_u^{(y)}-\vec{\ell}_u\rVert$ is the distance between the antenna element at SU and each REs on the RIS positioned at the near-field of the SU \cite{Han2020}, and $E_u=d_{u,x}d_{u,y}$.

The power of the reflected signal received by the RIS-s from $E_{u,n_u}$ can be represented as
\begin{align}
 P_{u,n_u}^r=\frac{G_uP_{u,n_u}^{in}|\rho_{u,n_u}|^2}{4\pi d_{n_s,n_u}^2}F_u^{rc}E_s,
\end{align}
where $F_u^{rc}=F_u^r(\theta_{u,n_u}^r,\varphi_{u,n_u}^r)F_u^{rx}(\theta_{u,n_u}^{rx},\varphi_{u,n_u}^{rx})$, $d_{n_s,n_u}$ is the distance between the $n_u^{\text{th}}$ RE on the RIS-u positioned at the near-field of the SU and the $n_s^{\text{th}}$ RE on the RIS-s positioned at the near-field of the SAP, and $E_s=d_{s,x}d_{s,y}$. Given the coordinates of RIS-s and RIS-u deployments, $d_{n_s,n_u}=\lVert\vec{\ell}_s^r+n_s^{(x)}d_{s,x}\vec{b}_s^{(x)}+n_s^{(y)}d_{s,y}\vec{b}_s^{(y)}-\vec{\ell}_u^r+n_u^{(x)}d_{u,x}\vec{b}_u^{(x)}+n_u^{(y)}d_{u,y}\vec{b}_u^{(y)} \rVert$.
The power of the reflected signal received by the SU from $E_{s,n_s}$ traversed through $E_{u,n_u}$ is given by
\begin{align}
 P_{n_s,n_u}^r=\frac{|\rho_{s,n_s}|^2G_sP_{u,n_u}^{r}}{4\pi d_{s,n_s}^2}F_s^{rc} A_r,
\end{align}
where $F_s^{rc}=F_s^r(\theta_{s,n_s}^r,\varphi_{s,n_s}^r)F_s^{rx}(\theta_{s,n_s}^{rx},\varphi_{s,n_s}^{rx})$, $d_{s,n_s}=\lVert \vec{\ell}_s-\vec{\ell}_s^r+n_s^{(x)}d_{s,x}\vec{b}_s^{(x)}+n_s^{(y)}d_{s,y}\vec{b}_s^{(y)}\rVert$ is the distance between the antenna element at the SAP and each REs on the RIS-s positioned at the near-field of the SAP \cite{Han2020}, and $A_r$ is the aperture of the receiving antenna at the SAP.
Consequently, the electric field of the reflected signal from $E_{s,n_s}$ traversed through $E_{u,n_u}$ received by the SAP can be expressed as
\begin{align}
 E_{n_s,n_u}^r=&=\sqrt{\frac{2Z_0P_{n_s,n_u}^r}{A_r}}e^{\left(\frac{-j2\pi d_{n_s,n_u}^c}{\lambda}+\phi_{n_u}+\phi_{n_s}\right)}\\
 &=\sqrt{\frac{Z_0P_tG_tG_sG_uE_sE_uF_{n_s,n_u}^c}{32\pi^3}}\nonumber\\
 &\quad \quad \frac{\rho_{s,n_s}\rho_{u,n_u} e^{\left(\frac{-j2\pi d_{n_s,n_u}^c}{\lambda}+\phi_{n_s}+\phi_{n_u}\right)}}{d_{u,n_u}d_{n_s,n_u}d_{s,n_s}},
\end{align}
where $Z_0$ is the characteristic impedance of the air, $d_{n_s,n_u}^c=d_{s,n_s}+d_{n_s,n_u}+d_{u,n_u}$, $E_s=d_{s,x}d_{s,y}$, and ${\left(\frac{-j2\pi d_{n_u,n_s}^c}{\lambda}+\phi_{n_s}+\phi_{n_u}\right)}$ is the phase alteration caused by the signal propagation and the reflections from the RIS-s and RIS-u. Moreover, the combined normalized radiation pattern is $F_{n_s,n_u}^c= F_u^{tc}F_u^{rc}F_s^{rc}$.
The total electric field of the received signal at the SAP can be defined as the superposition of the reflected electric fields by all REs at RIS-s towards the SAP, which is expressed as
\begin{align}\label{eq:ef}
 E(r)= \sum_{n_s \in \mc{N}_s} \sum_{n_u \in \mc{N}_u}E_{n_s,n_u}^r.
\end{align}
By knowing the received electric field at the SAP, its received signal power can be represented as
\begin{align}\label{eq:pr}
 P_r=\frac{|E(r)|^2}{2Z_0}A_r, \quad A_r = \frac{G_r\lambda^2}{4\pi}.
\end{align}
The maximum received power by substituting \eqref{ef} in \eqref{pr} and after the beamformer design at the RIS-s and RIS-u becomes
\begin{align}
 P_r=P_t\frac{G_tG_sG_uG_rE_sE_u\lambda^2a^2}{256\pi^4}\left|\underset{\substack{n_s \in \mc{N}_s\\n_u \in \mc{N}_u}}{\sum} \frac{\sqrt{F_{n_s,n_u}^c}}{d_{u,n_u}d_{n_s,n_u}d_{s,n_s}}\right|^2.\nonumber
\end{align}
The beamforming design of RIS-s and RIS-u is discussed in \secref{bfdesign}.
Thereby, the path loss of the double-RIS assisted near-field beamforming can be represented as

 \begin{align}\label{eq:plrl_nf}
 L_r^{nf}= \frac{256\pi^4}{G_tG_sG_uG_rE_sE_u\lambda^2a^2\left|\underset{\substack{n_s \in \mc{N}_s\\n_u \in \mc{N}_u}}{\sum} \frac{\sqrt{F_{n_s,n_u}^c}}{d_{u,n_u}d_{n_s,n_u}d_{s,n_s}}\right|^2}.
 \end{align}
Similarly, the path loss of the double-RIS assisted far-field beamforming can be represented as \cite{Tang}
\begin{align}\label{eq:plrl_ff}
 L_r^{ff}= \frac{256\pi^4d_u^2d_s^2d_h^2}{G_tG_sG_uG_rE_sE_uN_s^2N_u^2\lambda^2a^2F_{n_s,n_u}^c},
 \end{align}
 where $d_h$ is the distance between the centers of RIS-s and RIS-u.

\section{Performance Analysis of the Proposed System}\label{sec:PA}
This section investigates the impact of integrating two RISs on the overall performance of satellite communication links.
Some of the key aspects of performance analysis are as follows.
\subsection{Signal Enhancement and Dual-Stage Beamforming:}\label{sec:bfdesign}
{
The independent dual RIS deployment enables dual-stage beamforming, with the SU signal reflected by the first RIS near the SU to the second RIS near the satellite, without increased latency compared to single or no RIS systems. The second RIS then fine-tunes the beam to maximize the signal strength at the satellite's antenna.} This process significantly enhances the received signal power compared to a direct satellite link.

The channel links between the RIS-u to SU, $\vec{h}_u$, and the SAP to RIS-s, $\vec{h}_s$, follow the near-field channel model defined in \eqref{nfchannel}. Consequently, $\vec{h}_u$ and $\vec{h}_s$ can be expressed as
\begin{align}
 \vec{h}_u=\vec{b}(\theta_u,\varphi_u,d_u),~\text{and }
 \vec{h}_s=\vec{b}(\theta_s,\varphi_s,d_s),
\end{align}
respectively, where $d_u$ is the distance between the SU and the center of the RIS-u placed at its near-field and $d_s$ is the distance between the antenna element on the SAP and the center of the RIS-s placed at its near-field.
Conversely, the far-field channel representation in \eqref{ffchannel} can be used to simulate the channel between the RIS-s and RIS-u, $\vec{H}$.
As such, the signal that was received in \eqref{rxsig} can be rewritten as
\begin{align}\label{eq:rxdsignal}
 y &=\sqrt{\frac{P_t}{L_r}}\Big\{\vec{b}(\theta_s,\varphi_s,d_s)^T\vec{\Phi}_se^{-jkd_{h}}\vec{a}_{n_s}(\theta_{n_s},\varphi_{n_s})\nonumber\\
&\quad \quad\vec{a}_{n_u}^T(\theta_{n_u},\varphi_{n_u})\vec{\Phi}_u \vec{b}(\theta_u,\varphi_u,d_u)\Big\}+z.
\end{align}

Based on \eqref{rxdsignal}, the power of the received signal at the SU can be depicted as
\begin{align}
P_r&=\frac{P_t}{L_r}\Big|\vec{b}(\theta_s,\varphi_s,d_s)^T\vec{\Phi}_se^{-jkd_{h}}\vec{a}_{n_s}(\theta_{n_s},\varphi_{n_s})\nonumber\\
&\quad \quad\vec{a}_{n_u}^T(\theta_{n_u},\varphi_{n_u})\vec{\Phi}_u\ \vec{b}(\theta_u,\varphi_u,d_u)\Big|^2,
\end{align}
where $L_r$ is defined in \eqref{plrl_nf} and \eqref{plrl_ff} in accordance with the placement of RIS near the SU.
Now, in order to have maximum received power at the SAP, the beamformer at RIS-u and RIS-s can be configured as
\begin{align}
 \vec{\Phi}_u[n_u]&=\left(\vec{a}_{n_u}(\theta_{n_u},\varphi_{n_u})[n_u]\vec{h}_u[n_u]\right)^*, ~\text{and}\\
 \vec{\Phi}_s[n_s]&=\left(\vec{h}_s[n_s]\vec{a}_{n_s}(\theta_{n_s},\varphi_{n_s})[n_s]\right)^*e^{jk(d_{h})},
\end{align}
{respectively \cite{Han2020}}, where we consider normalized channel. Thereby, the maximum received SNR at the SAP can be expressed as
\begin{align}\label{eq:rsnr}
 \Gamma_r&=\frac{P_t}{L_rN_0}\left|\vec{h}_s^T\vec{\Phi}_s\vec{H}\vec{\Phi}_u\vec{h}_u\right|^2.
\end{align}

 \subsection{Achievable Data Rate}
  The achievable data rate (ADR) of a DRIS-assisted satellite communication system can be expressed using Shannon's capacity formula as follows,
  \begin{align}\label{eq:rate}
   R&= \log_2(1+\Gamma_r),
  \end{align}
which takes into account the SNR at the receiver, $\Gamma_r$, in \eqref{rsnr}. 
 	 \begin{figure*}
 	  \begin{subfigure}{0.33\textwidth}
      \centering
      \includegraphics[width=1.06\linewidth]{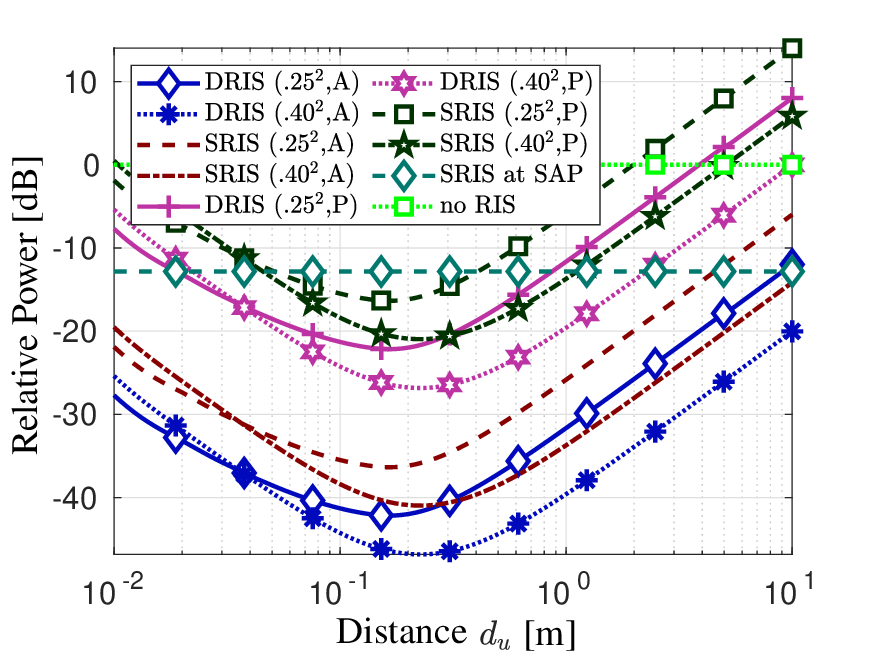}
      \caption{RP vs $d_u$, $\text{ADR}=2$ bps/Hz}
      \label{fig:prvsd_area}
 	  \end{subfigure}
 	   \begin{subfigure}{0.33\textwidth}
      \centering
      \includegraphics[width=1.06\linewidth]{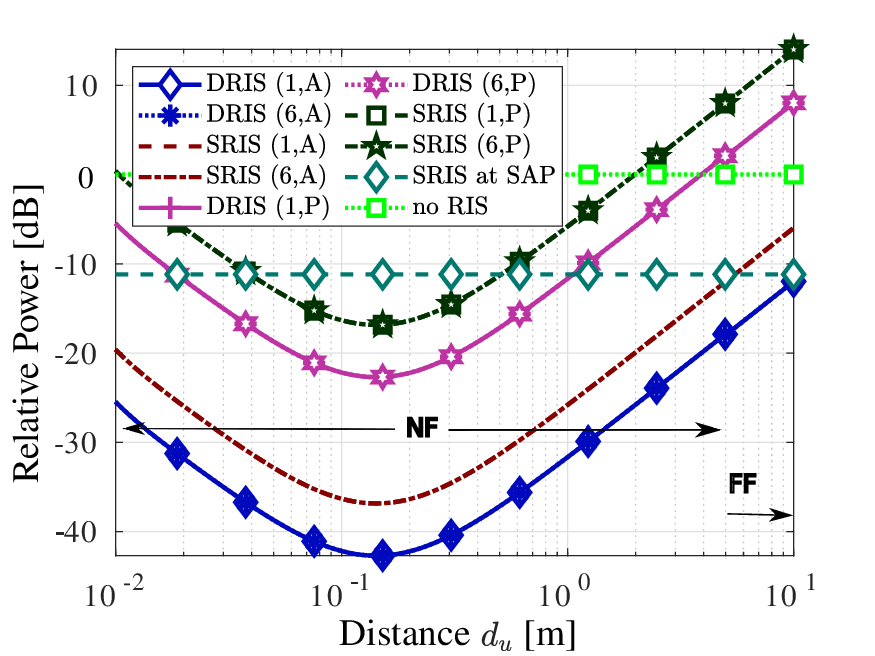}
      \caption{RP vs $d_u$, $A_u=0.25^2$ $\text{m}^2$}
      \label{fig:prvsd_rate}
 	  \end{subfigure}
      \begin{subfigure}{0.33\textwidth}
      \centering
      \includegraphics[width=1.06\linewidth]{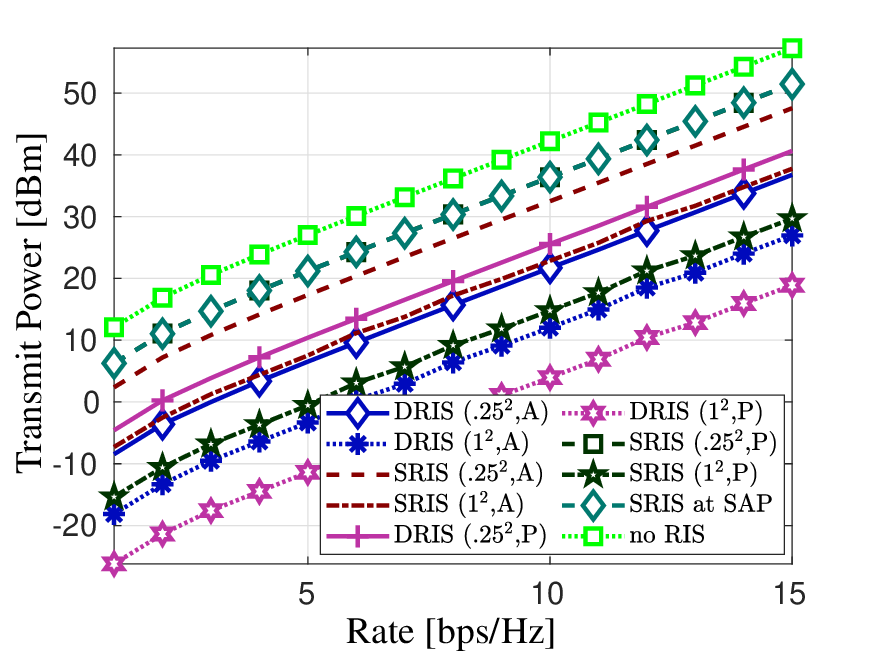}
      \caption{TP vs ADR, $d_u=1$ $\text{m}$}
      \label{fig:prvsrate_d}
      \end{subfigure}
      \caption{Relative power and transmit power required for various system parameters of double RIS, single RIS, and no RIS scenarios in the parenthesis as (a) `($A_u$, Active/Passive)' (b) `(ADR, Active/Passive)' (c) `($A_u$, Active/Passive)'.}
      \label{fig:fig2}
 	 \end{figure*}
  \subsection{Energy Efficiency}
 The energy efficiency (EE) of a DRIS-assisted satellite communications can be expressed as the proportion of the total ADR to the total power consumption.
 The total power consumption of the proposed system is determined as \cite{Sanila23}
 \begin{align}
  P_{\text{total}}=P_t+P_{\text{RIS s}}+P_{\text{RIS u}}+P_{\text{static}},
 \end{align}
 where $P_{\text{RIS s}}$ and $P_{\text{RIS u}}$ are the power consumed by the RIS near the SAP and SU, respectively, and $P_{\text{static}}$ is the power consumed by other components such as the satellite transceiver, control circuits, etc, which are static in nature. Consequently, the EE $\eta_{EE}$  is given by
 \begin{align}
  \eta_{EE}=\text{BW}\frac{R}{P_{\text{total}}}=\frac{\text{BW}\log_2\left(1+\Gamma_r\right)}{P_t+P_{\text{RIS s}}+P_{\text{RIS u}}+P_{\text{static}}},
 \end{align}
 where $\Gamma_r$ is defined in \eqref{rsnr} and BW is the bandwidth of the communication system.
The power consumed by a passive RIS $P_{RIS}$ consisting of $N$ REs is $NP_{ps}$, where $P_{ps}$ is the power consumed by each adaptive phase shifter circuitry. On the other hand, the power consumed by an active RIS becomes $P_{RIS}=P_A+NP_{dy}+P_{st}$, where $P_A$ is the maximum amplification power from the active RIS. Moreover, the dynamic and static power consumption of the active REs are denoted as $P_{dy}$ and $P_{st}$, respectively, with the note that $P_{ps}\ll P_{dy}$ \cite{Yigit23}.

\section{Simulation Results and Discussions}\label{sec:SR}
This section focuses on the performance comparison between DRIS-assisted mobile satellite communication systems and systems with a single RIS (SRIS) or no RIS. The `SRIS' scenario with simulation parameters in the parenthesis refers to the location of the RIS solely near the SU, whereas the `SRIS at SAP' configuration assumes that the RIS is placed only close to the SAP when the simulation parameters are not specified in the parenthesis. In contrast, the `no RIS' configuration indicates a satellite communication system that has only a direct link between the SAP and the SU.
 The simulation parameters are listed in \tabref{sp} \cite{Tekbiyik2022,Tang}.
 \begin{table}[htbp]
 	\centering
 	\caption{Simulation Parameters}\label{tab:sp}
 	\begin{tabular}{| p{5.5cm} | p{1.8cm} |}
 		\hline
 		\textbf{Parameters} & \textbf{Values} \\ \hline
 		The carrier frequency $f_c$ & $12$ GHz\\ \hline
 		The bandwidth  & 50 MHz\\ \hline
 		Distance between SAP and RIS-s $d_s$ & $1$ m \\ \hline
 		Distance between RIS-u and RIS-s $d_h$ & $1200$ km\\ \hline
Transmit power $P_t$& $10$ dBm \\ \hline
Noise variance $\sigma_s^2$, $\sigma_d^2$& $-133.5$ dBm\\ \hline
Dynamic power $P_{dy}$ & $30$ dBm\\ \hline
Static power of active RIS $P_{st}$& $35$ dBm\\ \hline
Static power of the system $P_{static}$& $75$ dBm\\ \hline
Phase-shift circuitry power $P_{ps}$& $5$ mW\\ \hline
Maximum amplification power at active RIS $P_{A}$& $\frac{P_t}{2}$ \\ \hline
Number of REs at RIS-s, $N_s$& $400$ \\ \hline
Gain at SU $G_t$ & $10$ dBi \\ \hline
Gain at SAP $G_r$ & $20$ dBi \\ \hline
Gain at RIS $G_s$, $G_u$& $3$ dBi \\ \hline
 	\end{tabular}
 	 \end{table}

\figref{prvsd_area} compares the relative power (RP) required at varying distances between the SU and its nearby RIS-u for a fixed ADR. In \figref{prvsd_rate}, the RP is shown for different ADRs with respect to the distance $d_u$ while maintaining a constant RIS-u area. Additionally, \figref{prvsrate_d} illustrates the impact of transmit power (TP) required across various ADRs and RIS-u areas, $A_u$, with the RIS-u at a fixed distance from the SU. The legends in \figref{prvsd_area} and \figref{prvsrate_d} follow the convention `System name ($A_u$, A/P)', where $A_u=N_ud_{u,x}d_{u,y}$ is the area of the RIS-u at the SU, which is measured in square meters, $\text{m}^2$. Additionally, `A', and `P' stand for active and passive deployment of RIS-u near the SU. The legends in \figref{prvsd_rate} are denoted as `System name (ADR, A/P)'. The path loss model dynamically transitions between the near-field and far-field models as in \eqref{plrl_nf} and \eqref{plrl_ff}, respectively, throughout the simulations based on the Rayleigh distance criterion as discussed in \secref{Channel}.

\begin{figure*}
 	 \begin{subfigure}{0.33\textwidth}
      \centering
      \includegraphics[width=1.06\linewidth]{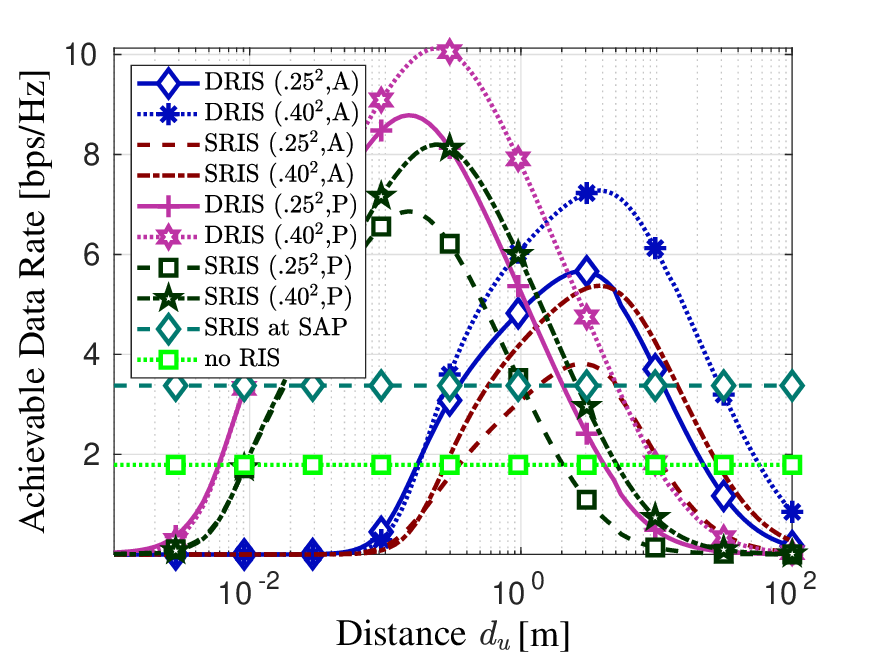}
      \caption{ADR vs distance $d_u$}
      \label{fig:rpvspt}
 	  \end{subfigure}
 	  \begin{subfigure}{0.33\textwidth}
      \centering
      \includegraphics[width=1.06\linewidth]{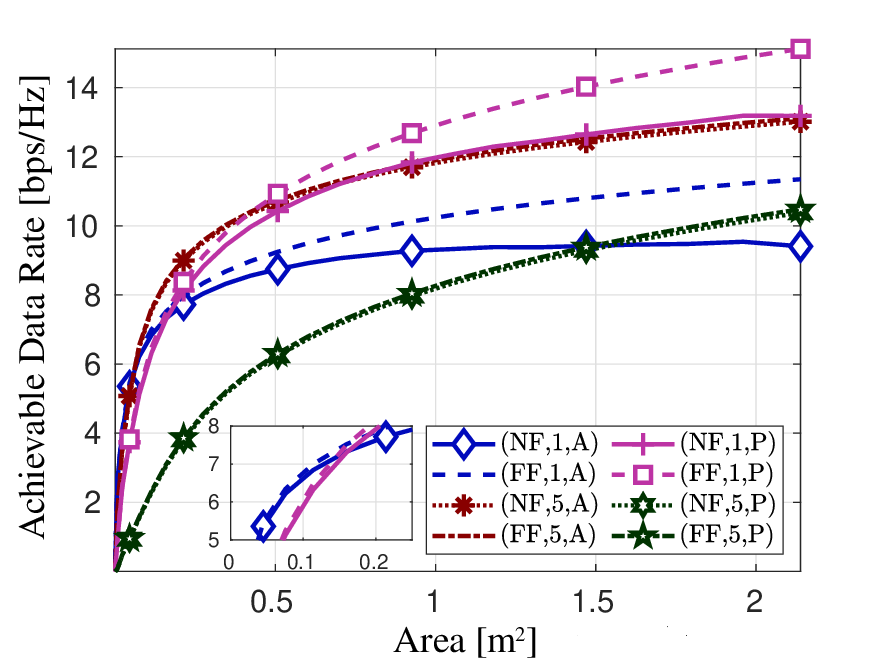}
      \caption{ADR vs area of RIS-u}
      \label{fig:arvsarea}
      \end{subfigure}
        \begin{subfigure}{0.33\textwidth}
      \centering
\includegraphics[width=1.06\linewidth]{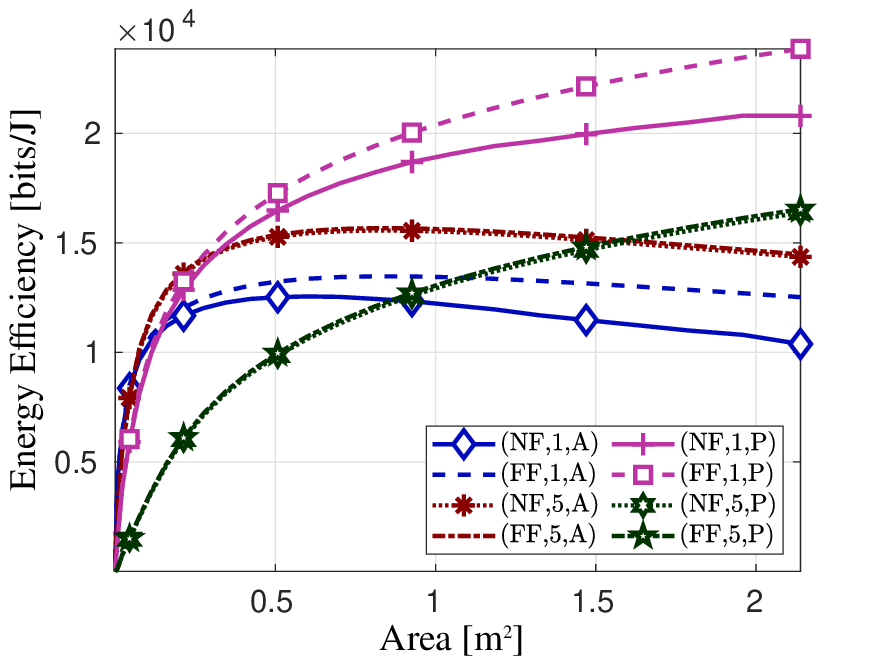}
\caption{EE vs area of RIS-u}
\label{fig:eevsarea}
 	  \end{subfigure}
\caption{(a) Performance analysis of double RIS, single RIS, and no RIS scenarios with system parameters as `($A_u$, Active/Passive)' (b,c) Performance analysis of the proposed DRIS system with system parameters as `(NF/FF, $d_u$, Active/Passive)'}
 	 \end{figure*}

\figref{rpvspt} presents a comparative analysis of the ADR for various distances $d_u$ with the legends follow the convention abstracted in \figref{prvsd_area}. On the other hand, \figref{arvsarea} and \figref{eevsarea} illustrate the ADR and EE comparison with different RIS-u areas at the SU, $A_u$. The legends of \figref{arvsarea} and \figref{eevsarea} adhere to the format `(NF/FF, $d_u$, A/P),' where `NF' and `FF' stand for near-field and far-field path loss models, as in \eqref{plrl_nf} and \eqref{plrl_ff}, respectively, which are used to obtain the ADR for the DRIS system. In contrast, the near-field and far-field path loss models of the SRIS system are defined as per \cite{Tang}. 

The insights from these results are elaborated as follows: From \figref{fig2}, it can be observed that as the SU moves closer to the RIS-u in the near field, the gain contribution from the farther REs progressively deteriorates due to the reduced field of view. Thereby, to keep the ADR constant, the SU transmits with more power.
Additionally, more RP is needed as the distance between the SU and RIS increases beyond the optimal $d_u$ distance. The reason for the rise in RP is that as distance increases in the far field, the RIS's effective gain decreases, requiring more power to compensate for the weaker signal for the intended ADR. However, when the RIS area expands, both active and passive RIS systems utilize less RP, as seen in \figref{prvsd_area}. Also, as \figref{prvsd_rate} illustrates, for both active and passive RIS implementations, the RP demand remains unchanged as the data rate increases. Passive RIS systems become more power-efficient as their surface area increases, as depicted in \figref{prvsrate_d}. The less power needed by the DRIS system (around $40$ dB less than the no RIS case) is due to its better signal-focusing ability. 

In \figref{rpvspt}, the ADR decreases significantly when the RIS is positioned very close to the SU, as the farther REs contribute less to the total gain, due to very small field of view. As the RIS distance from the SU increases, the ADR initially improves due to increased field of view and therefore stronger contributions from almost all REs but then declines in the far-field as signal strength diminishes over distance. Additionally, passive RIS performs better in the near-field, while active RIS outperforms at greater distances from the SU.
 
 The increase in RIS-u area always results in increased ADR for passive and active RIS systems, as seen in \figref{arvsarea}, by increasing the signal gain. When the RIS-u area increases, active RIS elements can magnify the reflected signal, resulting in higher ADR than with passive RIS elements. As opposed to this, passive RIS reflects the signal without amplifying it, which makes it more energy-efficient, particularly with larger RISs where more passive elements continue to enhance the signal without consuming more power.
 The EE advantages of active RIS decrease after reaching a particular RIS area since active RIS requires more power for the amplifier circuitry, as depicted in \figref{eevsarea}.
 Passive RIS systems with a larger surface area are preferred where power efficiency is critical. However, active RIS might be more appropriate for systems with less concern for power and needing maximum signal enhancement when only a smaller RIS area is feasible.

\section{Conclusion}\label{sec:conclusion}
We proposed a low-power double RIS-assisted mobile LEO satellite communication architecture, with one RIS positioned near the satellite antenna and another near the user’s antenna. This design demonstrates minimized power consumption and maximized data rates, enabling highly energy-efficient LEO satellite communications for mobile users, even with very small RISs, whether active or passive. Notably, the double RIS setup achieves a $40$ dB reduction in required power compared to the no-RIS scenario, using an RIS near the user with an area of just $0.25^2 \text{ m}^2$. 
Investigating scenarios in which several mobile devices are spatially scattered in the near-field region of the same RIS while the RIS is concurrently communicating with a satellite is an enticing direction for future research.

\bibliographystyle{ieeetr}
\bibliography{references}
\end{document}